\documentclass[twocolumn,showpacs,preprintnumbers,amsmath,amssymb]{revtex4-1}
\usepackage[latin1]{inputenc}
\usepackage[francais]{babel}

\usepackage{graphicx,epsf,subfigure}  

\begin{document}
 
\title{Modeling Washboard Road:\\ from experimental measurements to linear stability analysis}
 
\author{Baptiste Percier}
\affiliation{Universit\'e de Lyon, Laboratoire de Physique, \'Ecole Normale Sup\'erieure de Lyon, \\
CNRS UMR 5672, 46 All\'ee d'Italie, 69364 Lyon cedex 07, France. }
\author{S\'ebastien Manneville}
\affiliation{Universit\'e de Lyon, Laboratoire de Physique, \'Ecole Normale Sup\'erieure de Lyon, \\
CNRS UMR 5672, 46 All\'ee d'Italie, 69364 Lyon cedex 07, France. }
\affiliation{Institut Universitaire de France}
\author{Nicolas Taberlet}
\affiliation{Universit\'e de Lyon, UFR de Physique, Universit\'e Claude Bernard Lyon 1, and Laboratoire de Physique, \'Ecole Normale Sup\'erieure de Lyon, \\
CNRS UMR 5672, 46 All\'ee d'Italie, 69364 Lyon cedex 07, France. }

\date{\today}

\begin{abstract}

When submitted to the repeated passages of vehicles unpaved roads made of sand or gravel can develop a ripply pattern known as washboard or corrugated road.
We propose a stability analysis based on experimental measurements of the force acting on a blade (or plow) dragged on a circular sand track and show that a linear model is sufficient to describe the instability near onset.
The relation between the trajectory of the plow and the profile of the sand bed left after its passage is studied experimentally. The various terms in the expression of the lift force created by the flow of granular material on the plow are determined up to first order by imposing a sinusoidal trajectory to the blade on an initially flat sand bed, as well as by imposing a horizontal trajectory on an initially rippled sand bed. Our model recovers all the previously observed features of washboard road and accurately predicts the most unstable wavelength near onset as well as the critical velocity for the instability.

\end{abstract}
\maketitle

\section{Introduction}

When submitted to the repeated passages of vehicles, gravel and sand roads can develop a ripply pattern known as washboard or corrugated road. Not only is the bumpiness of the track a disturbance to drivers, but it also causes a loss of adherence and control and is therefore a real hazard. A similar phenomenon occurs on train, tramway or metro tracks and is known as rail corrugation. Due to its obvious economic importance, rail corrugation, which is caused by wear or plastic deformation, has been extensively studied as reported in the engineering literature~\cite{sato_2002} through experiments~\cite{Bellette_2011}, field work~\cite{grassie_2008} and theoretical and numerical analysis~\cite{Bellette_2010,Hempelmann_1995,Sandstrom_2012,Wu_2011,Xie_2008,Xie_2008,Meehan2008}. The formation of a washboard road over a sand bed was first studied experimentally in~\cite{mather63}, and theoretical models have been proposed in~\cite{both01washboard,Kurtze2001} but include an ad-hoc diffusion term whose physical meaning remains unclear. Recently, it has been shown that a washboard road can develop when a simple inclined blade or plow (instead of a rolling wheel) is dragged along a sand road~\cite{Taberlet2007,Bitbol2009}. The formation of an instability during a unique passage of a plow under its own weight over an initially flat surface has also been studied in viscous or viscoplastic fluids~\cite{Hewitt2012a,Hewitt2012}.

The aim of the present paper is to derive a model for the washboard instability caused by the repeated passages of a plow over a sand bed, from experimental measurements of the forces acting on the plow. Previous work has focused on force measurements in the simpler case of a blade at constant altitude steadily dragging a mound of sand over a flat sand bed~\cite{Percier_2011}. Herein, we extend these results by using force measurements to probe the mechanical response to an oscillatory excitation. These measurements are then used as the basis for a linear stability analysis of the washboard road instability.

The paper is organised as follows:
Section~\ref{sec:previous} gives a list of previous experimental observations and results that a model should recover. Section~\ref{sec:experimental} presents the experimental methods while section~\ref{sec:framework} introduces the framework and assumptions of the linear stability analysis. The relation between the trajectory of the plow and the shape left in the sand bed after its passage is discussed in section~\ref{sec:reshaping}. The expression of the lift forces acting on the plow is discussed in section~\ref{sec:lift}. Finally the predictions of the linear stability analysis are compared to experimental results in section~\ref{sec:stability}.

\section{\label{sec:previous}Previous Results}

Our previous work has established a number of characteristic features which a model for washboard road ought to reproduce. (i)~There exists a critical velocity, $v_c$, below which the sand bed remains flat when perturbed and above which any irregularity will develop into a regular rippled pattern. 
(ii)~This critical velocity increases with increasing mass of the plow or wheel, following a power law. 
(iii)~For velocities greater than the critical velocity the rippled washboard road pattern appears only gradually over several passages of the plow. While the initial wavelength, $\lambda$ keeps a finite and well-defined value, the amplitude of the ripples grows continuously from zero to a saturated value.

In the present paper we focus on the onset of the instability, i.e. on velocities greater yet close to the critical velocity. The saturation and coarsening of the pattern remain to be studied in further detail.

Previous work has shown that in the case of washboard road caused by a plow, compaction of the sand bed has not been observed nor seems to play any role~\cite{Bitbol2009}, contrary to the case of a rotating wheel. This may be due to the fact that when using a plow the granular material is constantly shuffled at every passage of the plow. Therefore compaction will be neglected throughout the present stability analysis. Note that this assumption would be highly debatable had a wheel been used.

For the range of velocities and masses of the plow used we found that the ''horizontal'' length of the plowed material, $L_0$ (typically a few centimeters), remains smaller than the wavelength of the washboard road pattern, $\lambda$ typically ten centimeters. Its influence is briefly discussed in section~\ref{subsec:corrugated} but unless otherwise mentioned it will be neglected.

\begin{figure}[htbp]
\begin{center}
{\includegraphics*{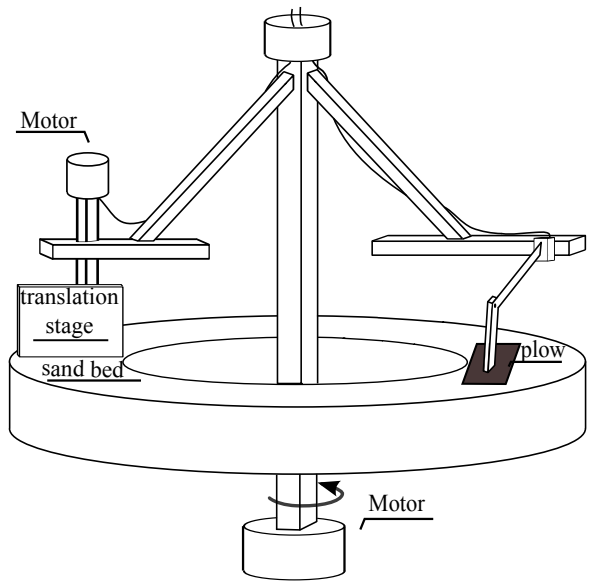}}
\caption{Sketch of the experimental setup. A plow is dragged at a constant horizontal velocity on a 5-m-long circular track filled with sand.}\label{fig0}
\end{center}
\end{figure}

\section{\label{sec:experimental}Experimental setup}

\begin{figure}[htbp]
\begin{center}
{\includegraphics*{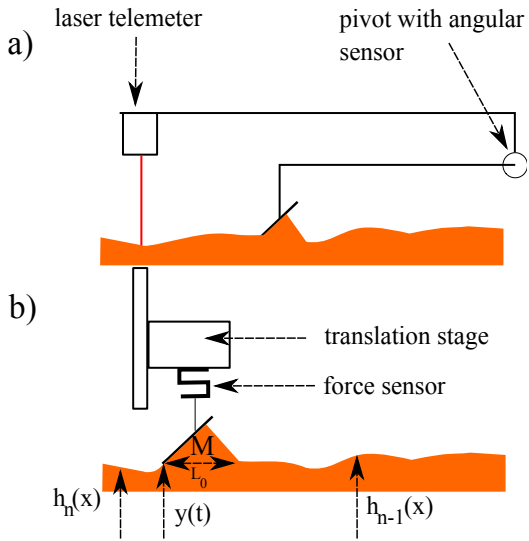}}
\caption{Sketch of the experimental setups. a) Plow attached to an articulated arm and free to move vertically. b) Plow mounted on force sensors attached to a computer-controlled translation stage which allows one to impose the trajectory.}\label{fig2}
\end{center}
\end{figure}

The experimental setup consists in a circular track of length 5~m (average diameter of 80 cm), width 25~cm, filled with a 20-cm high layer of granular material (figure~\ref{fig0}). The granular material used is sand-blasting sand of size ranging from 200 $\rm \mu m$ to 400 $\rm  \mu m$.
The plow consisting of a flat blade made in PVC (inclined at $45^{\circ}$) of width 14.5 cm is dragged at a constant horizontal velocity $v$, ranging from 0.1 $\rm m\, s^{-1}$ to 2 $\rm m\, s^{-1}$. These values of the speed are high enough to produce a continuous-flow regime and low enough to avoid a gaseous regime.
The plow is either attached to an arm whose end is free to rotate (figure~\ref{fig2}a) or mounted on a translation stage (figure~\ref{fig2}b).

In the former case (figure~\ref{fig2}a), the length of the arm is 50 cm while the amplitude of the ripples near onset is typically of the order of a few millimeter. Therefore, we will consider that the arm remains nearly horizontal and the pendulum motion of the arm is neglected, so that any change in either the horizontal speed or in the inclination of the plow due to a change in altitude is neglected. The vertical dynamics of the plow is therefore governed only by its own weight and the lift force caused by the plowed material, while its horizontal motion is imposed by a motor and the washboard road pattern may develop. Compared to most engineering work~\cite{sato_2002} our system is simplified in that the plow has no tire nor suspension. Movies in the supplementary material show both the steady-state washboard instability on real-time and the growth of the ripples through stroboscopic images.  
The vertical position of the plow is recorded using a magnetic angle sensor (ASM-PRAS1) placed on one end of the arm.
Unless otherwise mentioned, prior to any experiment the sand bed is made flat by dragging (over several tens of rotations) a gradually rising  vertical blade around the track.

In the second case (figure~\ref{fig2}b) the plow is rigidly mounted on a vertical translation stage ($\rm 5~\mu m$ accuracy), while its horizontal velocity remains imposed. Two force sensors (Testwell KD40S) are used to measure the vertical lift force acting on the plow. In this second case the trajectory of the plow is imposed through a NIUSB6259 card. The sand bed profiles prior to and after the passage of the plow are measured using two laser telemeters (optoNCDT 1302
from Micro-Epsilon not shown on figure~\ref{fig2}b) of 0.02 mm accuracy. Obviously no washboard road pattern can appear using this setup used only to probe the mechanical response of the material to an imposed trajectory.


The following variables are defined: $x$ is the horizontal position of the lower tip of the plow (set to zero at time $t=0$), $y(x= v (t-(n-1)T))$ its vertical position where $n$ is the number of passages and $T$ the duration of one passage, $h_n(x)$ is the profile of the sand bed after the $n^{th}$ passage of the plow ($h_0$ being therefore the initial profile prior to the experiment) (see figure~\ref{fig2}).

Note that since the horizontal velocity is imposed there always exists a correspondence between the position $x$ and the time $t$. Any variable can therefore be differentiated with respect to $x$ or $t$, although for convenience $h_n$ will be expressed as a function of $x$ and $y$ as a function of $t$.

The sand bed being largely thicker than the amplitude of the ripples near onset it can be considered infinitely deep (having halved the thickness of the sand bed in our experiments showed no noticeable effects). The reference of the vertical variable $y$ and $h_{n}$ is therefore arbitrary.

\section{\label{sec:framework}Framework of the linear stability analysis}

The washboard road pattern emerges from the interaction between the plow and the sand bed. The trajectory of the plow is governed by Newton's second law where the vertical lift force, $f_l$, acting on the plow may depend on the plowed mass $M$, the vertical position of the plow $y$, and the sand bed profile $h$ and their derivatives. 
Similarly the shape of the sand bed depends on the trajectory of the plow, while the new profile left after the $n^{th}$ passage, $h_n(x)$, may depend on the lift, the previous profile $h_{n-1}(x)$ and the trajectory of the plow $y(x)$.

\begin{figure}[htbp]
\begin{center}
\includegraphics*{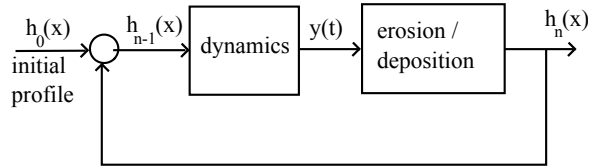}
\caption{Diagram representing the coupling between the dynamics of the plow and the reshaping of the sand bed due to erosion and deposition. }\label{fig3}
\end{center}
\end{figure}

The plow position $y$ is governed by its dynamics (weight and lift force) while an erosion/deposition law reshapes the bed profile $h$ (compaction being neglected). Although these two processes are coupled and simultaneous a convenient way to represent the coupling is the schematics of figure~\ref{fig3} commonly used in the engineering literature~\cite{Bellette_2010,Hempelmann_1995,Meehan2008}. In the following section the reshaping of the sand bed is discussed whereas the dynamics of the plow is studied in section~\ref{sec:lift}.

As mentioned in the introduction the present paper proposes a linear stability analysis near the onset of the washboard road instability. 
It will be shown that although nonlinearities may be needed to account for the saturation and coarsening of the ripple pattern, a linear model accurately reproduces the experimental data.
The linear response is probed using sinusoidal excitations (of either the vertical position of the plow, or of the initial bed thickness) of wave-number $k$, or corresponding angular velocity $\omega =vk$.
As mentioned above, the origin of the vertical axes is arbitrary and does not play any role. Since compaction is here neglected the average height and thickness profile is constant and can be set to zero. Therefore, the system can be described using complex variables indicated by an under bar. $\underline{M}(t) = \langle M \rangle + \underline{M}_0 \, e^{i \omega t}$, where $\langle M \rangle$ is the plowed mass on time-average, and similarly $\underline{y} = \underline{A}_y \, e^{i \omega t}$ and $\underline{h} = \underline{A}_h \, e^{i k x}$.
As a reminder, based on previous experimental observations the four following assumptions are made.
(i) A 2-dimensional model is suitable, (ii) compaction is neglected and (iii) the length of the plowed material ($L_0$ on figure~\ref{fig2}) is smaller than the wavelength of the pattern and (iv) near the onset of the instability studied here the plow remains in contact with the sand bed.

\section{\label{sec:reshaping}Reshaping of the sand bed}

In rail corrugation, the time evolution of the bed profile can involve wear or plastic deformation~\cite{sato_2002} and is governed by the normal and tangential forces acting between the wheel and the track, as well as by the trajectory of the wheel $y(t)$ and the previous shape of the track, $h_{n-1}(x)$, these three variables being coupled. Here in the case of a plow running over a sand bed the sand is simply eroded and redeposited (compaction being neglected).

\begin{figure}[htbp]
\begin{center}
{\includegraphics*{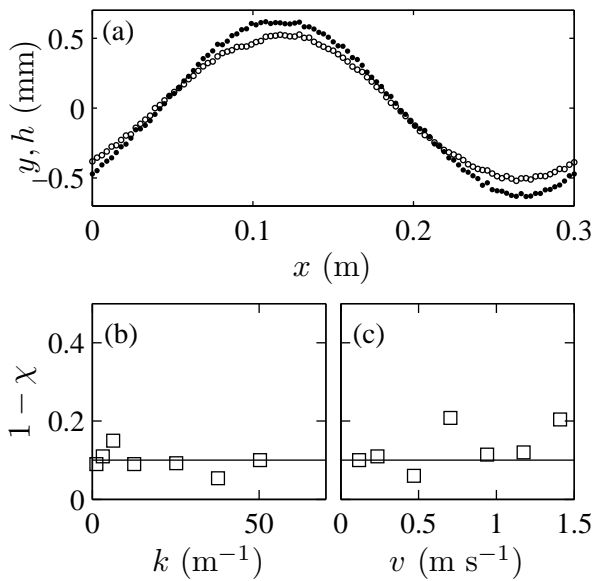}}
\caption{ a) Plots of the trajectory of the plow $y$ (solid symbols) and the profile left after the passage of the plow, $h_n$ (open symbols), for $k=21~{\rm m^{-1}}$ and $v=0.5~{\rm m\,s^{-1}}$.
The profile $h_n$ measured after the passage of the plow is translated by the distance to the tip of the plow. 
b) and c) Ratio between the amplitude of the profile $h_n$ and the trajectory, $ \chi = h_n / y$ as a function of $k$ (for $v=0.5~{\rm m\,s^{-1}}$) and $v$ (for $k=21~{\rm m^{-1}}$). }
\label{fig4}
\end{center}
\end{figure}

In order to investigate the relation between $h_n(x)$ and the other variables the plow is dragged at constant horizontal velocity $v$ while a vertical sinusoidal  trajectory of angular velocity, $\omega$, (and corresponding wavenumber $k=\omega / v$) is imposed to the plow mounted on the translation stage: $\underline{y}(t) = A_y e^{i \omega t}$. The profile, $h_n(x)$ is then recorded. Figure~\ref{fig4} shows a typical example of the outcome for $v=0.5$ $\rm m\,s^{-1}$, $A_y=0.6$ mm and $k=21$ m$^{-1}$ (corresponding wavelength $\lambda = 30$ cm).

The first result is that there is no phase-lag between the imposed vertical position, $y$, and the profile left after the passage of the plow, $h_n$ (see Figure~\ref{fig4}a).
Our experiments have shown that this remains true for the range of values of $v$, $A_y$ and $k$ of interest in this paper (near the onset of the instability, $v$ is ranging from 0.2 to $2.0~{\rm m\,s^{-1}}$ and $k$ is ranging from 10 to 125 $\rm m^{-1}$).

The major result is that the amplitude of $h_n$ is slightly smaller than that of $y$, meaning that the plow leaves a footprint in the sand which is similar yet not identical to its trajectory.
The ratio between the two amplitudes, $\chi$, is plotted as a function of wavenumber and velocity (in figure~\ref{fig4} b and c). The ratio $\chi$ is clearly independent of the wavenumber  and depends very weakly on the velocity. The ratio $\chi$ can be seen as an imprinting efficiency ratio. It seems intuitive that it should tend toward one for low velocities since when trying to carve a shape in a sand bed, one should be as delicate (and slow) as possible. Note however that over the range of velocities of interest $\chi$ remains rather constant at $\chi \approx 0.9$ and any dependence on $v$ will be neglected in the following.
Moreover, the ratio $\chi$ is also found to be independent of the amplitude of the imposed trajectory, $A_y$, for the range of small amplitudes studied ($A_y < 5~{\rm mm}$).
In the present case where the trajectory is imposed by the translation stage the profile left after the $n^{th}$ passage of the plow, $h_n$, is found to be independent of that prior to this passage, $h_{n-1}$. 
In the following, we will therefore consider that the profile left after the $n^{th}$ passage of the plow, $h_n(x)$ is simply given by:

\begin{equation}
h_n(x) =  \chi \; y(x)
\label{eq_chi}
\end{equation}

\noindent where $\chi =0.9 \pm 0.05$. Although this value is close to unity it will emerge that the difference between the trajectory of the plow and the shape left in the sand bed is a key element of the washboard road instability.

\section{\label{sec:lift}Dynamics of the plow}

\subsection{Jerk equation}

In a situation where the plow is free to move vertically its dynamics is simply governed by the lift force acting on the plow $f_l$ (its horizontal position being imposed):


\begin{equation}
m \ddot{y} = -m g + f_l
\label{eq_dyn}
\end{equation}

Our recent study~\cite{Percier_2011} has shown that in a uniform motion over a flat bed (constant horizontal velocity and no vertical motion), the lift force is simply proportional to the plowed mass $M$ and is  independent of the velocity: $f_{l} = \mu M g$, where $\mu$ is an effective solid friction coefficient which depends solely on the inclination of the plow (or angle of attack), and $\mu = 0.56$ for an inclination of $45^{\circ}$. The goal of this section is to extend our previous analysis to non-uniform motion over a rippled sand bed. 

From mass conservation (and having neglected the length $L_0$, see figure~\ref{fig2}) the rate at which mass is gained or lost is simply given by:

\begin{equation}
\dot{M} = \rho v \left( h_{n-1}(x) - h_n(x)  \right)  
\label{eq_Mdot}
\end{equation}

\noindent where $\rho$ is the 2D density of the sand ($\rho$ = bulk density of sand $\times$ packing fraction $\times$ plow width).

In a steady state (horizontal or sinusoidal motion) equations (\ref{eq_dyn}) and  (\ref{eq_Mdot}) simply give a relation between the average plowed mass and the mass of the plow:

\begin{equation}
\mu \langle M \rangle = m
\label{eq_Mmoy}
\end{equation}

On an initially flat sand bed ($h_{n-1}(x)=0$) differentiating equation (\ref{eq_dyn}) and using equations (\ref{eq_chi}) and (\ref{eq_Mdot}) yields:

\begin{equation}
 h_n^{'''}(x) \; + \; K^3 \; h_n(x) \; = \; 0 
\label{eq_jerk}
\end{equation}

\noindent where $K = (\mu g \rho \chi / (m v^2) )^{1/3}$. 
This equation known as a jerk equation governs the dynamics of third-order oscillators and charged particules in motion in their own electromagnetic field.
Equation (\ref{eq_jerk}) has oscillatory solutions (of wavenumber $\sqrt{3}K/2 $) but with an exponentially growing amplitude along the track during one unique passage over a flat bed, meaning that the road is always unstable. This is in total contradiction with the experimental observation of a critical velocity, at least in the range of masses and velocities under study. Therefore additional dissipative terms are needed in equation (\ref{eq_jerk}). Note however that the wavelength predicted by the jerk equation (\ref{eq_jerk}) is rather close to that observed near the onset of the instability (i.e. the most unstable mode for a velocity close to the critical velocity). Indeed, for $v=0.8~{\rm m\,s^{-1}}$ and $m=0.25$ kg, equation~(\ref{eq_jerk}) predicts a wavelength $\lambda=\frac{4\pi}{\sqrt{3}K}=0.41~{\rm m}$ close to the experimental wavelength of the pattern for similar values of the parameters~\cite{Taberlet2007,Bitbol2009}.

To the first order described in this section, the lift force depends on the plowed mass $M$, i.e. on $\int h_n$ and $\int h_{n-1}$ as seen from equation~\ref{eq_Mdot}. 
The simplest mathematical form is to include the derivatives of $h_n$ and $h_{n-1}$, whose physical meaning will be discussed below:
$f_l = f(\int h_n, \int h_{n-1}, h_n, h_{n-1}, h_n', h_{n-1}')$. Under the linearity assumption used here it will be assumed that the effects of all further terms are additive. Section~\ref{subsec:flat} is devoted to the study of the case $h_{n-1}=0$ whereas section~\ref{subsec:corrugated} will focus on the case where $h_n=0$ and $\underline{h}_{n-1}(x) = A_h e^{ikx}$.

\subsection{\label{subsec:flat}Lift force over an initially flat bed}

This section aims at determining the dependence of the lift force on $h_n$ and $h_{n}'$ in the case of an initially flat and horizontal bed, $h_{n-1}=0$.  The plow is mounted on the translation stage and the vertical force is recorded as the vertical position is imposed.

An initial mass of sand $\langle M \rangle $ is plowed by lowering the plow in the sand bed. From this position (at $t=0$) a sinusoidal motion is imposed to the plow: $\underline{y} = A_y e^{i \omega t } = A_y e^{i k x }$. The sand bed is flattened at every rotation using a vertical blade attached to another arm which erases any pre-existing profile. This allows the data to be averaged over typically 30 oscillation periods. 
The lift force reads:

\begin{equation}
f_l(t)  = \mu M  g - b \, y - c \, \dot{y} 
\label{eq_lift_1}
\end{equation}

\noindent where $b$ and $c$ are real and positive coefficients whose dependence on the average plowed mass $\langle M \rangle$ and on the velocity is key to the stability analysis. Note that in the steady regime studied here equation (\ref{eq_lift_1}) could be expressed as a function of $h_n$ and $h_n'$ using equation (\ref{eq_chi}).

Although the two additional terms $by$ and $c\dot{y}$ are introduced as the simplest linear extension of the lift force, their physical meaning is easily understood. The first additional term can be seen as a restoring force acting on an intruder, adding an extra force which is proportional to the penetration depth $y$. Clearly when pushing an intruder into the sand bed ($y<0$), the additional force is positive, hence the negative sign in $-b y$. Such linear restoring forces have been reported in the literature~\cite{Schiffer_2009,Peng2009,Seguin_2011,STONE}.
The second term may be seen as a contribution of the vertical penetration speed (or equivalently of the local slope of the profile when $h_{n-1} \neq 0$). Again the additional force should be positive when penetrating into the sand bed ($\dot{y}<0$), hence the negative sign in $-c \dot{y}$.

In order to measure the coefficients $b$ and $c$ and to study their dependence on the average mass $\langle M \rangle$ and velocity $v$, the transfer function $\underline H$ is defined: $\underline H = \underline f_l / \underline y$.
Equation (\ref{eq_lift_1}) yields the following band-stop filter expression:

\begin{equation}
\underline H = - b + i \left( \frac{\mu g \rho \chi v}{\omega} - c \, \omega \right)
\label{eq_H}
\end{equation}

\begin{figure}[htbp]
\begin{center}
{\includegraphics*{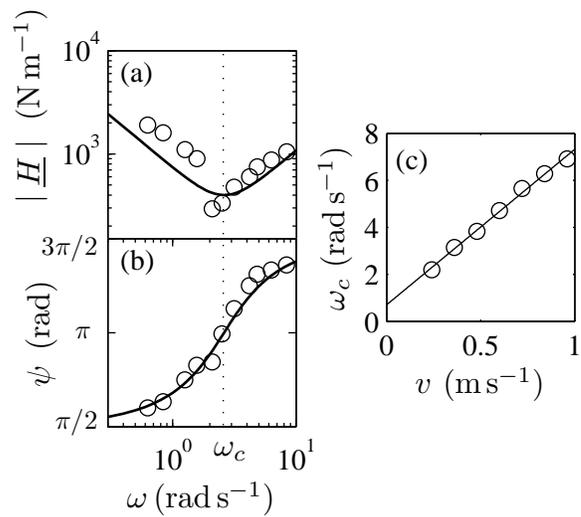}}
\caption{a) Absolute value and argument of the transfer function $| \underline H |$ for $v=0.3~{\rm m\,s^{-1}}$ and $\langle M \rangle$ = 0.58 kg. The solid lines correspond to equation~(\ref{eq_H}) the dashed line corresponds to $\omega_c$ for which $\underline H$ is real and $| \underline H|$ is minimal. b) $\omega_c$ as a function of $v$ showing a linear dependence for $\langle M \rangle$ = 0.58 kg.}
\label{fig5}
\end{center}
\end{figure}

Equation (\ref{eq_H}) predicts that there exists a minimum in $| \underline H |$ when $\underline H$ is real and negative ($\psi = \arg(\underline H) = \pi$), i.e. for $\omega = \omega_c \equiv \left(     \mu g \rho \chi v /c   \right) ^{1/2} $.  Therefore the coefficient $c$ can be computed from the value of $\omega_c$ while the coefficient $b$ is simply the minimum reached by $| \underline H |$. In the framework of equation (\ref{eq_lift_1}) measuring $\min( | \underline H | )$ and $\omega_c$ is sufficient to experimentally determine the values of the two coefficients $b$ and $c$. The experimental data of the absolute value has a higher noise level (about $15\%$) than the argument (about $5\%$) and hence, the value of $\omega_c$ was measured from the argument $\psi(\omega_c) = \pi$ (dashed line on figure~\ref{fig5}).

The transfer function $ \underline{H}$ was measured for various plowed masses, $\langle M \rangle$, and velocities, $v$, and a typical example is shown on Figure~\ref{fig5} ($v=0.3~{\rm m\,s^{-1}}$ and $\langle M \rangle$ = 0.58 kg). The experimental data shows the features predicted by equation (\ref{eq_H}) (minimum in $| \underline H |$, reached for $\underline H$ real ($\psi = \pi$), $\psi$ ranging from $\pi/2$ to $3\pi/2$) and both the absolute value and the argument are well fitted by equation (\ref{eq_H}) (solid lines).
For the small amplitudes imposed here (typically $A_y <~$ 5~mm) we found no dependence on the excitation amplitude $A_y$. These results validate the linear form of equation (\ref{eq_lift_1}).

\begin{figure}[htbp]
\begin{center}
{\includegraphics*{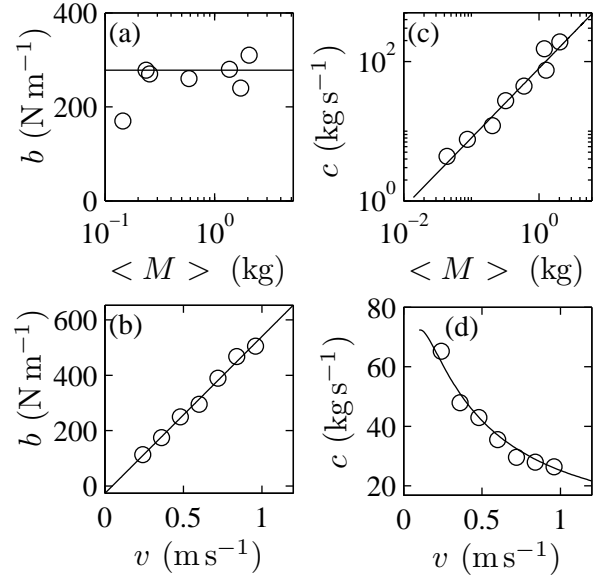}}
\caption{ (a),(c), Coefficients of the additional terms $b$ and $c$ as a function of $\langle M \rangle$ (for $v=0.5~{\rm m\,s^{-1}}$) and $v$ (for $\langle M \rangle =1$~kg). $b$ appears to be proportional to $v$ (solid line of slope one) and independent of $M$. $c$ is proportional to $M$ and its dependence on $v$ is given by the expression of $\omega_c$ found in figure~\ref{fig5}c (solid line). }
\label{fig6}
\end{center}
\end{figure}

The values of $\omega_c$ computed from the transfer function are plotted in figure~\ref{fig5}c as a function of $v$ for an averaged plowed mass $\langle M \rangle =0.58$ kg. It appears that $\omega_c$ is a linear function of the velocity $v$: $\omega_c \propto v + v_0$, where $v_0 = 0.1 \pm 0.01$ $\rm m\,s^{-1} $. The physical meaning of this dependence remains to be explained. Still this empirical expression along with measurements of $\min (| \underline H |)$ for various plowed masses and velocities allows to plot the two coefficients $b$ and $c$ as functions of $\langle M \rangle$ and $v$ with a typical uncertainty of 10\%  (figure~\ref{fig6}). The following observations emerge: over the range of parameters of interest, $b$ does not show any systematic dependence on $\langle M \rangle$ and is a linear function of $v$ (Fig.~\ref{fig6}a and ~\ref{fig6}b) and the following expression will be used $b = B_0 \, v$, with $B_0 = 560$ $\rm kg\,m$$^{-1}$.s$^{-1}$. The coefficient $c$ appears to be proportional to the mass $\langle M \rangle$ (solid line of slope one in figure~\ref{fig5}c) whereas its dependence on $v$ has already been determined from the empirical expression of $\omega_c$ (solid line on figure~\ref{fig5}). Overall the following expression emerges:
$ c = C_0 \mu \langle M  \rangle g \rho \chi  v / (v+v_0)^2 $, where $C_0$ is a constant $C_0=8.7~10^{-2}$ m.kg$^{-1}$.

\subsection{\label{subsec:corrugated}Lift force over a rippled sand bed}

The previous section has established empirical expressions for the contribution of the derivatives of $y(t)$ (or equivalently $h_n(x)$) to the lift force, $f_l$. In this section the contribution to $f_l$ of a pre-existing profile prior to the $n^{th}$ passage of the plow is studied by shaping a sinusoidal profile ($\underline h_{n-1} = A_h e^{ikx} = A_h e^{i \omega t}$) over which a horizontal trajectory is imposed ($\underline y=0$, and hence $\underline h_n=0$) with a blade plowing an average mass of sand $\langle M \rangle$.

Similar additional terms to the lift forces are expected although with an opposite sign. Indeed a bump in the sand bed ($h_{n-1} > 0$) will create an extra positive force as will a positive slope $h'_{n-1}$:

\begin{equation}
f_l  =  \mu  M  g  + \tilde b \, h_{n-1}  + \tilde c \, \dot{h}_{n-1}
\label{eq_lift_2}
\end{equation}

\noindent where $\tilde b$ and $\tilde c$ and coefficients whose dependence on the averaged plowed mass $\langle M \rangle$ and on the velocity $v$ has to be determined.

\begin{figure}[htbp]
\begin{center}
{\includegraphics*{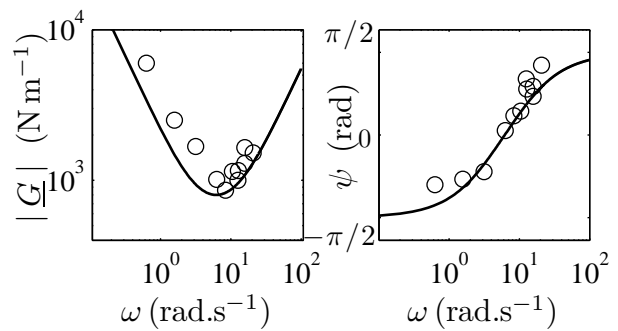}}
\caption{Absolute value and argument of the transfer function $| \underline G |$ for $M=1$ kg and $v=1.5~{\rm m\,s^{-1}}$. The solid lines correspond to equation~(\ref{eq_lift_2}) with parameters inferred from section~\ref{subsec:flat}, showing good agreement with the experimental data.}
\label{fig7}
\end{center}
\end{figure}

It is expected that the role of $y$ in equation (\ref{eq_lift_1}) and $h_{n-1}$ in equation (\ref{eq_lift_2}) are symmetrical and we propose that $\tilde b = b$ and $\tilde c= c$. A new transfer function can be defined as $\underline G = \underline f_l / \underline h_{n-1} $ and is plotted as a function of $\omega$ on figure~\ref{fig7} (for $\langle M \rangle =1~{\rm kg}$ and $v=1.5~{\rm m\,s^{-1}}$, one can note that $v$ is high compared to the previous experiment, this is to prevent the effect of the length $L_0$ and make sure that $\lambda \gg L_0$). The solid lines correspond to the predictions of equation (\ref{eq_lift_2}) using the expressions of $b$ and $c$ obtained in section~\ref{subsec:flat}. They show a good agreement with the experimental data and validate the proposed expression of the lift force.

\section{\label{sec:stability}Stability analysis}

\subsection{Exponential growth rate}

Having empirically established the mass and velocity dependence of additional terms in the lift force in section~\ref{sec:lift} using a plow whose trajectory is imposed, the equation of motion of a plow free to move vertically can be given by combining equations (\ref{eq_dyn}),(\ref{eq_lift_1}) and (\ref{eq_lift_2}) under the assumption of additive forces made above:

\begin{equation}
m \ddot y = -m g + \mu g M(t) + b(v) \, (h_{n-1} - y ) + c(\langle M \rangle,v) ( \dot h_{n-1} - \dot y ) 
\label{eq_dyn2}
\end{equation}

Differentiating equation (\ref{eq_dyn2}) and using equation (\ref{eq_chi}) the growth rate of the instablitiy, $\sigma= \ln  \frac{|h_n|}{|h_{n-1}|} $, can be expressed:

\begin{equation}
\sigma = \frac{1}{2} \, \left( \ln \frac{ (\alpha-\gamma k^2)^2 + (\beta k)^2 }{ (\alpha \chi -\gamma k^2)^2 + (\beta k - k^3)^2} \right)  + \ln \chi 
\label{eq_growth}
\end{equation}

\noindent where $\alpha = \mu g \rho / (mv^2) $, $\beta = B_0 / (m v)$ and $\gamma  = C_0 g \rho \chi /   (v+v_0)^2$.



%
%
%
%
%

\begin{figure}[htbp]
\begin{center}
{\includegraphics*{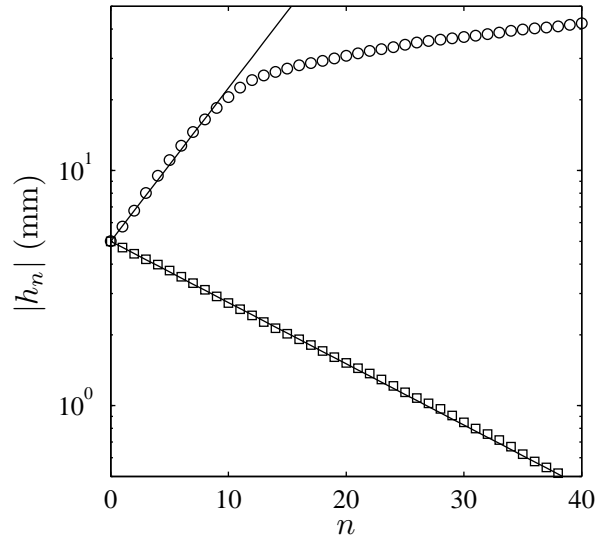}}
\caption{\label{fig8a} $h_n$ as a function of the number of rotations for $v=1.1~{\rm m\,s^{-1}}$ and $\lambda=195~{\rm mm}$ ($\square$) and $v=0.8~{\rm m\,s^{-1}}$ and $\lambda=216~mm$ ($\circ$). Straight lines are exponential fits of the data over the first ten rotations.}

\end{center}
\end{figure}

Experimentally this growth rate is measured by preparing an initially rippled pattern (similarly to section~\ref{subsec:corrugated}) and instead of imposing a vertical trajectory to the plow it is left free to move vertically. After several rotations of the plow the ripples either vanish or increase depending on the wavelength of the pattern and on the velocity of the plow. Figure~\ref{fig8a} shows two typical examples of these behaviors. During the first rotations (at least 10), the amplitude of the ripples can be fitted to an exponential curve and the growth rate is calculated from the exponental fit.

\begin{figure}[htbp]
\begin{center}
{\includegraphics*{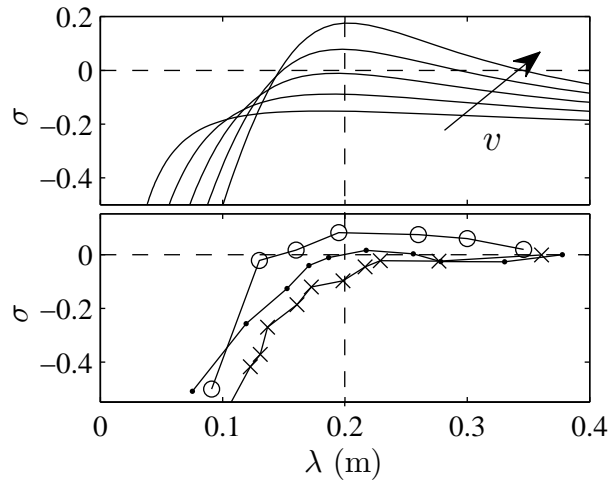}}
\caption{ (Top) Predictions of equation , growth rate $\sigma$ as a function of $\lambda$ for velocities ranging from 0.4 $\rm m\,s^{-1}$ to 1.2 $\rm m\,s^{-1}$ from bottom to top: (\ref{eq_growth}). (Bottom) Experimental measurements for $v=0.6~{\rm m\,s^{-1}}$ ($\times$), $v=0.8~{\rm m\,s^{-1}}$ ($ \bullet$) and $v=1.1~{\rm m\,s^{-1}}$ ($ \circ $). Above a critical velocity $v_c$ there exist positive values of $\sigma$ corresponding to the onset of the instability.}
\label{fig8}
\end{center}
\end{figure}

Using the values of the coefficients $B_0$ and $C_0$ determined experimentally in the previous sections, the  growth rate $\sigma$ can be computed from equation~\ref{eq_growth} as a function of the wavenumber $k$ (or corresponding wavelength $\lambda$) for various values of the average mass $\langle M \rangle$ and velocity $v$. Figure~\ref{fig8} (top) shows plots of $\sigma$ for $\langle M \rangle = 0.3$ kg and for velocities ranging from 0.4 $\rm m\,s^{-1}$ to 1.2 $\rm m\,s^{-1}$. For comparison, the experimental growth rate per rotationis shwon in figure~\ref{fig8} bottom.

The predictions of the model are in good qualitative and quantitative agreement with experimental results and show an important feature: for low velocities, the growth rate remains negative for all wavelengths, meaning that any perturbation will be gradually eroded, whereas for high velocities there exists a positive maximum corresponding to the fastest growing mode or the most unstable wavelength.

\subsection{Critical velocity and fastest growing mode}

The first velocity for which there exists a positive value  of the growth rate, $\sigma$, is the critical velocity of the instability, $v_c$.
The model derived not only recovers the existence of a critical velocity but also predicts the fastest growing mode: the theoretical and experimental most unstable wavelengths are in excellent agreement as shown in figure~\ref{fig8}.

It should be noted that there is some degree of uncertainty as to the asymptotic behavior of $\sigma$ since the expressions for the additional terms in equation~(\ref{eq_lift_1}) were determined over a limited range of wavelengths and velocities near the transition.

\begin{figure}[htbp]
\begin{center}
{\includegraphics*{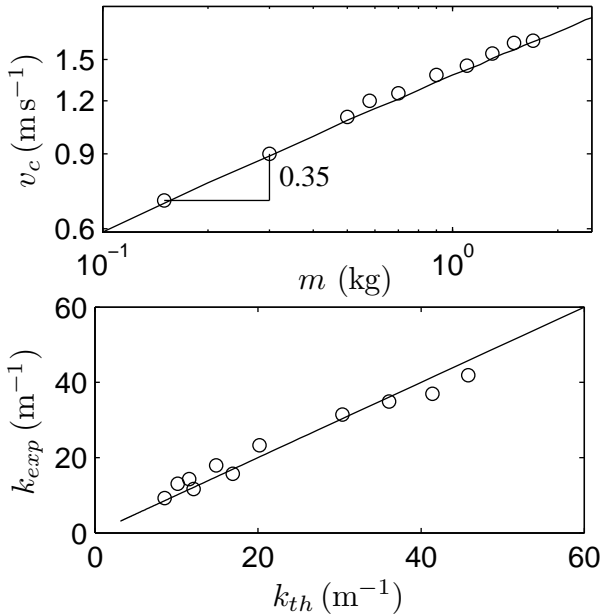}}
\caption{(Top)  Critical velocity $v_c$ as a function of the mass of the plow $m$ ($\circ$), the straight line corresponds to the prediction of the model and follows a power law with an exponent 0.35. (Bottom) Experimental wavenumber as a function of the predicted one. The straight line shows $k_{exp}=k_{th}$.  }
\label{fig9}
\end{center}
\end{figure}

The critical velocity $v_c$ for which the first positive value of $\sigma$ is observed depends only on one parameter: the average plowed mass $\langle M \rangle$, i.e. the mass of the plow $m = \mu \langle M \rangle$. Figure~\ref{fig9} (top) shows the values of $v_c$ inferred from our model for a plow mass ranging from 100 g to 2 kg. The critical velocity follows a power-law as a function of this mass: $v_c \propto m^{0.35}$, as found in previous experiments~\cite{Bitbol2009}, yet with a lower exponent ($v_c \propto m^{0.25}$). Although it is satisfactory to recover a power-law, the origin of the discrepancy in the exponent remains unclear. However the relevance of a power-law over only one decade is debatable. Again expanding the predictions of equation~(\ref{eq_growth}) to a wider range of masses, $m$, would be questionable since the expression of the lift force was determined in a somewhat narrow range of parameters near the onset of the instability.

Finally the value of the predicted most unstable wavenumber, $k_{th}$, can be compared to that measured experimentally, $k_{exp}$. The latter was not measured from the plots of $\sigma$ but instead was simply deduced from the profile of the sand bed after a few rotations. These values of $k_{th}$ and $k_{exp}$ depend on the mass of the plow $m$ as well as on the velocity, $v$ (clearly for $v > v_c(m)$). For simplicity, they are plotted not as a function of $m$ and $v$ but instead $k_{exp}$ is plotted as a function of $k_{th}$ on figure~\ref{fig9} (bottom). The agreement between the predicted and experimentally measured most unstable wavenumber is excellent for all values of the mass and velocity as shown by the solid line of slope one.

\section{Conclusion}

We have presented a linear stability analysis based on experimental measurements of the lift force acting on a blade plowing a mound of sand on a sand bed.
From previous experimental work~\cite{Percier_2011} which focused on the case of a steady plow over a flat surface
an equation for the dynamics of the plow was derived. However we found that further terms were needed to recover the observed features of the instability. The empirical expressions of these additional terms were obtained by probing the mechanical response of the system to sinusoidal excitations, both in the trajectory of the plow and in the initial profile of the sand bed. From this the growth rate of a sinusoidal perturbation was calculated and we showed that above a critical velocity, which depends on the mass of the plow, it displays a positive maximum corresponding to the most unstable or fastest growing mode. The critical velocity as well as the most unstable wavenumber were computed and showed excellent agreement with experimental measurements.

The expression of the lift force was obtained for a range of parameters near the onset of the washboard road instability. Hence the model recovers all the experimental observations near the threshold of the instability but fails to include the saturation of the pattern observed in experiments. Further measurements would be necessary and a thorough modelling will require to include nonlinearities.

Finally it would be interesting to extend the present study to the case of a rolling wheel. 
However this remains challenging since on the one hand the transport mechanism is clearly different and may involve plastic deformation and on the other hand compaction is expected to play a crucial role.

\section{Acknowledgement}
The authors would like to thank T. Divoux, V. Grenard, C. Perge, S.W. Morris, J.N.McElwaine and B. Andreotti for fruitful discussion. This work was supported by the F\'ed\'eration Amp\`ere de Physique, Universite Lyon 1.
%

%

\end{document}